\documentclass[nofootinbib,aps,prd,twocolumn,showpacs,preprintnumbers,amsmath,amssymb]{revtex4}
\usepackage{epsfig}

  \newcommand{\tr}{{\rm Tr}}
\def\lsim{\raise0.3ex\hbox{$<$\kern-0.75em\raise-1.1ex\hbox{$\sim$}}}
\def\gsim{\raise0.3ex\hbox{$>$\kern-0.75em\raise-1.1ex\hbox{$\sim$}}}

\begin{document}

\title{Static quark anti-quark interactions in zero and finite temperature QCD. \\
  I. Heavy quark free energies, running coupling and quarkonium binding}

\author{Olaf Kaczmarek} \email{okacz@physik.uni-bielefeld.de}
\affiliation{ Fakult\"{a}t f\"{u}r Physik, Universit\"{a}t Bielefeld, D-33615
  Bielefeld, Germany }

\author{Felix Zantow} \email{zantow@quark.phy.bnl.gov} \affiliation{Physics
  Department, Brookhaven Natl. Laboratory, Upton, New York 11973, USA}

\date{\today} 

\preprint{BI-TP 2005/08 and BNL-NT-05/8 }


\begin{abstract}
  We analyze heavy quark free energies in $2$-flavor QCD at finite temperature
  and the corresponding heavy quark potential at zero temperature. Static quark
  anti-quark sources in color singlet, octet and color averaged channels are
  used to probe thermal modifications of the medium. The temperature dependence
  of the running coupling, $\alpha_{qq}(r,T)$, is analyzed at short and large
  distances and is compared to zero temperature as well as quenched
  calculations. In parts we also compare our results to recent findings in
  $3$-flavor QCD. We find that the characteristic length scale below which the
  running coupling shows almost no temperature dependence is almost twice as
  large as the Debye screening radius. Our analysis supports recent findings
  which suggest that $\chi_c$ and $\psi\prime$ are suppressed already at the
  (pseudo-) critical temperature and thus give a probe for quark gluon plasma
  production in heavy ion collision experiments, while $J/\psi$ may survive the
  transition and will dissolve at higher temperatures.
\end{abstract}

\maketitle

\section{Introduction}
The study of the fundamental forces between quarks and gluons is an essential
key to the understanding of QCD and the occurrence of different phases which
are expected to show up when going from low to high temperatures ($T$) and/or
baryon number densities. For instance, at small or vanishing
temperatures quarks and gluons get confined by the strong force while at
high temperatures asymptotic freedom suggests a quite different QCD medium
consisting of rather weakly coupled quarks and gluons, the so-called quark
gluon plasma (QGP) \cite{PLP}. On quite general grounds it is therefore
expected that the interactions get modified by temperature. For the analysis of
these modifications of the strong forces the change in free energy due to the
presence of a static quark anti-quark pair separated by a distance $r$ in a
QCD-like thermal heat bath has often been used since the early work
\cite{McLerran:1980pk,McLerran:1981pb}. In fact, the static quark anti-quark
free energy which is obtained from Polyakov loop correlation functions
calculated at finite temperature plays a similar important role in the
discussion of properties of the strong force as the static quark potential does
at zero temperature.

The properties of this observable (at $T=0$: potential, at $T\neq0$: free
energy) at short and intermediate distances ($rT\;\lsim\;1$) is important for
the understanding of in-medium modifications of heavy quark bound states. A
quantitative analysis of heavy quark free energies becomes of considerable
importance for the discussion of possible signals for the quark gluon plasma
formation in heavy ion collision experiments \cite{Matsui:1986dk,RR}. For
instance, recent studies of heavy quarkonium systems within potential models
use the quark anti-quark free energy to define an appropriate finite
temperature potential which is considered in the non-relativistic Schr\"odinger
equation \cite{Digal:2001iu,Digal:2001ue,Wong:2001kn,Wong:2001uu}. Such
calculations, however, do not quite match the results of direct lattice
calculations of the quarkonium dissociation temperatures which have been
obtained so far only for the pure gauge theory
\cite{Datta:2003ww,Asakawa:2003re}. It was pointed out \cite{Kaczmarek:2002mc}
that the free energy ($F$) of a static quark anti-quark pair can be separated
into two contributions, the internal energy ($U$) and the entropy ($S$). The
separation of the entropy contribution from the free energy, {\em i.e.} the
variable $U=F+TS$, could define an appropriate effective potential at finite
temperature\footnote{
While a definition of the quark anti-quark potential can be given properly at
zero temperature using large Wilson loops, at finite temperature a definition
of the thermal modification of an appropriate potential energy between the
quark anti-quark pair is complicated \cite{Karsch:2005ex}.} \cite{Kaczmarek:2002mc,Zantow:2003ui}, $V_{\text{eff}}(r,T)\equiv
U$, to be used as
input in model calculations and might explain in
parts the quantitative differences found when comparing solutions of the
Schr\"odinger equation with direct calculations of spectral functions
\cite{Datta:2003ww,Asakawa:2003re}.  First calculations which use the internal
energy obtained in our calculations
\cite{Shuryak:2004tx,Wong:2004kn,Brown:2004qi,Park:2005nv} support this
expectation. Most of these studies consider so far quenched QCD. Using
potentials from the quenched theory, however, will describe the interaction of
a heavy quark anti-quark pair in a thermal medium made up of gluons only. It is
then important to understand how these results might change for the case of a
thermal heat bath which also contains dynamical quarks.

On the other hand, it is the large distance property of the heavy quark
interaction which is important for our understanding of the bulk properties of
the QCD plasma phase, {\em e.g.} the screening property of the quark gluon
plasma \cite{Kaczmarek:1999mm,Kaczmarek:2004gv}, the equation of state
\cite{Beinlich:1997ia,Karsch:2000ps} and the order parameter (Polyakov loop)
\cite{Kaczmarek:2003ph,Kaczmarek:2002mc,Dumitru:2004gd,Dumitru:2003hp}. In all
of these studies deviations from perturbative calculations and the ideal gas
behavior are expected and were indeed found at temperatures which are only
moderately larger than the deconfinement temperature. This calls for
quantitative non-perturbative calculations. Also in this case most of todays
discussions of the bulk thermodynamic properties of the QGP and its apparent
deviations from the ideal gas behavior rely on results obtained in lattice
studies of the pure gauge theory, although several qualitative differences are
to be expected when taking into account the influence of dynamical fermions;
for instance, the phase transition in full QCD will appear as an crossover
rather than a 'true' phase transition with related singularities in
thermodynamic observables. Moreover, in contrast to a steadily increasing
confinement interaction in the quenched QCD theory, in full QCD the strong
interaction below deconfinement will show a qualitatively different behavior at
large quark anti-quark separations. Due to the possibility of pair creation the
stringlike interaction between the two test quarks can break leading to a
constant potential and/or free energy already at temperatures below
deconfinement \cite{DeTar:1998qa}.

Thus it is quite important to extend our recently developed concepts for the
analysis of the quark anti-quark free energies and internal energies in pure
gauge theory \cite{Kaczmarek:2002mc,Kaczmarek:2003dp,Kaczmarek:2004gv,Phd} to
the more complex case of QCD with dynamical quarks, and to quantify the
qualitative differences which will show up between pure gauge theories and QCD.

\begin{table}[htbp]
\centering
\setlength{\tabcolsep}{0.7pc}
\begin{tabular}{|lll|lll|}
\hline
$\beta$ & $T/T_c$ & \# conf. & 
$\beta$ & $T/T_c$ & \# conf. \\
\hline
3.52  & 0.76    &  2000   &
3.72  & 1.16    &  2000   \\
3.55  & 0.81    &  3000   &
3.75  & 1.23    &  1000   \\
3.58  & 0.87    &  3500  &
3.80  & 1.36    &  1000   \\
3.60  & 0.90    &  2000   &
3.85  & 1.50    &  1000   \\
3.63  & 0.96    &  3000   &
3.90  & 1.65    &  1000   \\
3.65  & 1.00    &  4000   &
3.95  & 1.81    &  1000   \\
3.66  & 1.02    &  4000   &
4.00  & 1.98    &  4000   \\
3.68  & 1.07    &  3600  &
4.43  & 4.01    &  1600   \\
3.70  & 1.11    &  2000  &
      &         &        \\
\hline
\end{tabular}
\caption{Sample sizes at each $\beta$ value and the temperature in units of the
  (pseudo-) critical temperature $T_c$.}
\smallskip
\label{tab:configs}
\end{table}

\begin{figure}[tbp]
  \epsfig{file=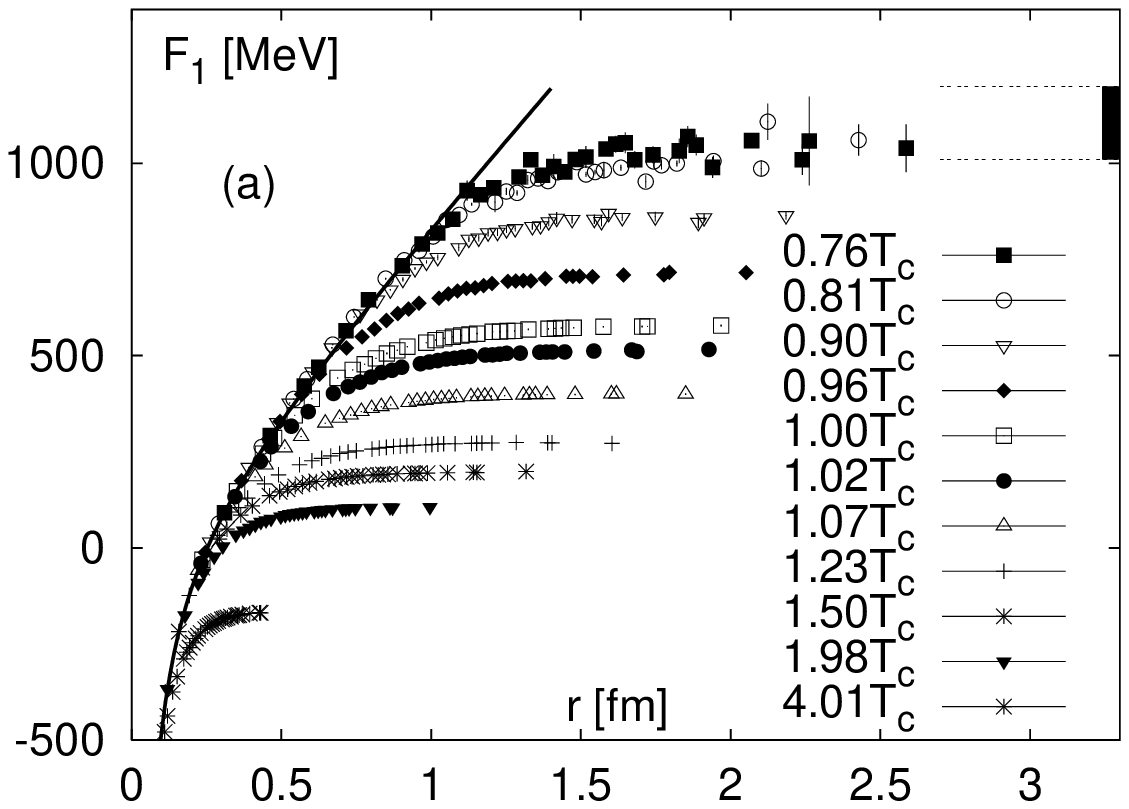,width=8.65cm}
  \epsfig{file=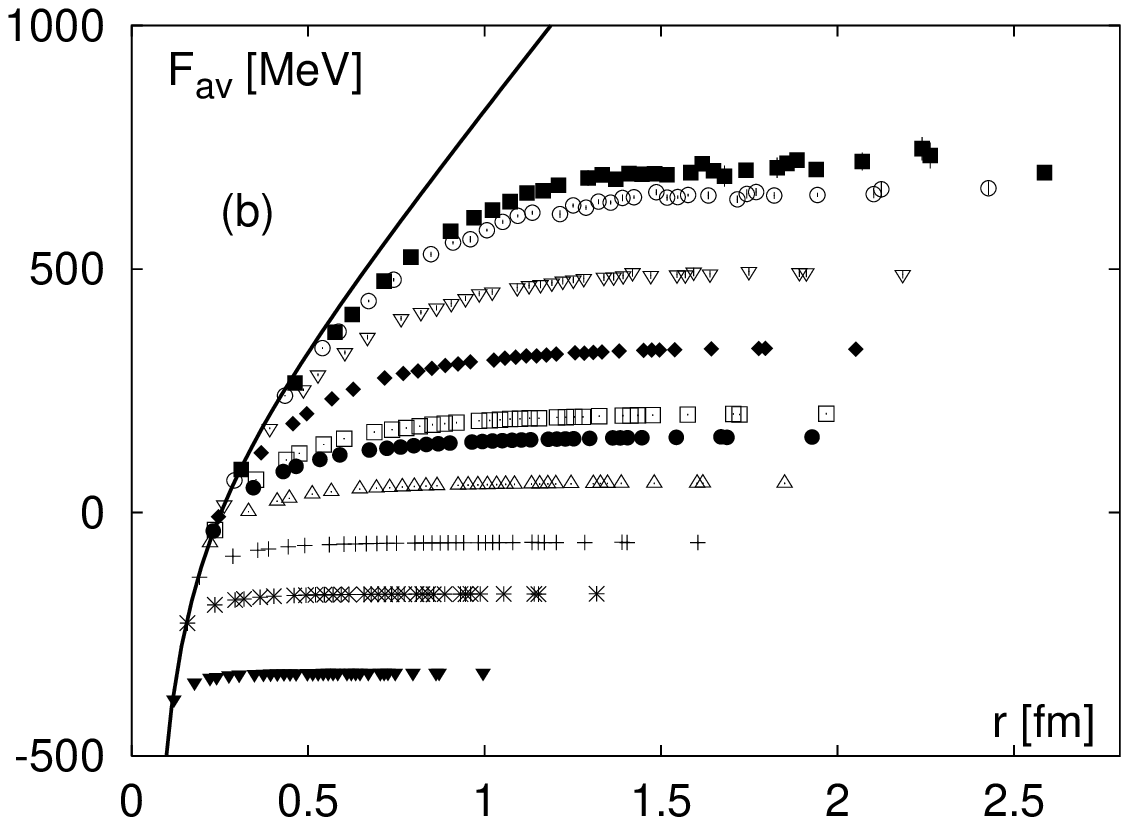,width=8.65cm}
\caption{
  (a) The color singlet quark anti-quark free energies, $F_1(r,T)$, at several
  temperatures close to the phase transition as function of distance in
  physical units. Shown are results from lattice studies of $2$-flavor QCD. The
  solid line represents in each figure the $T=0$ heavy quark potential, $V(r)$.
  The dashed error band corresponds to the string breaking energy at zero
  temperature, $V(r_{\text{breaking}})\simeq1000-1200$ MeV, based on the
  estimate of the string breaking distance, $r_{\text{breaking}}\simeq1.2-1.4$
  fm \cite{Pennanen:2000yk}. (b) The color averaged free energy, $F_{\bar q
    q}(r,T)$, normalized such that $F_{av}(r,T)\equiv F_{\bar q q}(r,T)-T\ln9$
  \cite{Kaczmarek:2002mc} approaches the heavy quark potential, $V(r)$ (line),
  at the smallest distance available on the lattice. The symbols are chosen as
  in (a).  }
\label{fes}
\end{figure}
For our study of the strong interaction in terms of the quark anti-quark free
energies in full QCD lattice configurations were generated for $2$-flavor QCD
($N_f$=2) on $16^3\times 4$ lattices with bare quark mass $ma$=0.1, {\em i.e.}
$m/T$=0.4, corresponding to a ratio of pion to rho masses ($m_{\pi}/m_{\rho}$)
at the (pseudo-) critical temperature of about $0.7$ ($a$ denotes the lattice
spacing) \cite{Karsch:2000kv}. We have used Symanzik improved gauge and
p4-improved staggered fermion actions. This combination of lattice actions is
known to reduce the lattice cut-off effects in Polyakov loop correlation
functions at small quark anti-quark separations seen as an improved restoration
of the broken rotational symmetry. For any further details of the simulations
with these actions see \cite{Allton:2002zi,Allton:2003vx}.  In
Table~\ref{tab:configs} we summarize our simulation parameters, {\em i.e.} the
lattice coupling $\beta$, the temperature $T/T_c$ in units of the pseudo
critical temperature and the number of configurations used at each
$\beta$-value. The pseudo critical coupling for this action is
$\beta_c=3.649(2)$ \cite{Allton:2002zi}. To set the physical scale we use the
string tension, $\sigma a^2$, measured in units of the lattice spacing,
obtained from the large distance behavior of the heavy quark potential
calculated from smeared Wilson loops at zero temperature \cite{Karsch:2000kv}.
This is also used to define the temperature scale and $a\sqrt{\sigma}$ is used
for setting the scale for the free energies and the physical distances.  For
the conversion to physical units, $\sqrt{\sigma}=420$MeV is used.  For
instance, we get $T_c=202(4)$ MeV calculated from $T_c/\sqrt{\sigma}=0.48(1)$
\cite{Karsch:2000kv}. In parts of our analysis of the quark anti-quark free
energies we are also interested in the flavor and finite quark mass dependence.
For this reason we also compare our $2$-flavor QCD results to the todays
available recent findings in quenched ($N_f$=0)
\cite{Kaczmarek:2002mc,Kaczmarek:2004gv} and $3$-flavor QCD
($m_\pi/m_\rho\simeq0.4$ \cite{Peterpriv}) \cite{Petreczky:2004pz}. Here we use
$T_c=270$ MeV for quenched and $T_c=193$ MeV \cite{Petreczky:2004pz} for the
$3$-flavor case.

Our results for the color singlet quark anti-quark free energies, $F_1$,
and color averaged free energies, $F_{av}$, are summarized in
Fig.~\ref{fes} as function of distance at several temperatures close to the
transition. At distances much smaller than the inverse temperature ($rT\ll1$)
the dominant scale is set by distance and the QCD running coupling will be
controlled by the distance. In this limit the thermal modification of the
strong interaction will become negligible and the finite temperature free
energy will be given by the zero temperature heavy quark potential (solid
line). With increasing quark anti-quark separation, however, thermal effects
will dominate the behavior of the finite temperature free energies ($rT\gg1$).
Qualitative and quantitative differences between quark anti-quark free energy
and internal energy will appear and clarify the important role of the entropy
contribution still present in free energies. The quark anti-quark internal
energy will provide a different look on the inter-quark interaction and thermal
modifications of the finite temperature quark anti-quark potential. Further
details of these modifications on the quark anti-quark free and internal
energies will be discussed.

This paper is organized as follows: We start in section~\ref{sect0} with a
discussion of the zero temperature heavy quark potential and the coupling. Both
will be calculated from $2$-flavor lattice QCD simulations. We analyze in
section~\ref{secfreee} the thermal modifications on the quark anti-quark free
energies and discuss quarkonium binding. Section~\ref{seccon} contains our
summary and conclusions.  A detailed discussion of the quark anti-quark
internal energy and entropy will be given separately \cite{pap2}.

\section{The zero temperature heavy quark potential and coupling}\label{sect0}

\begin{figure}[tbp]
  \epsfig{file=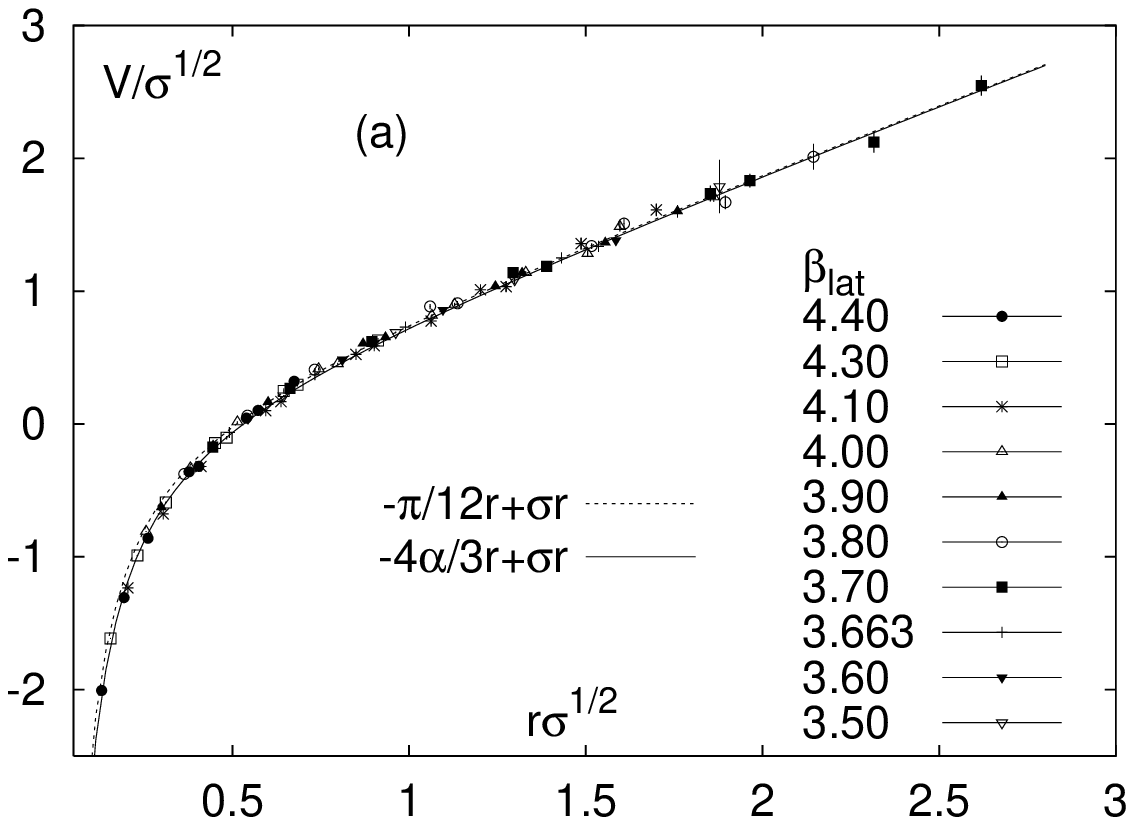,width=8.65cm}
  \epsfig{file=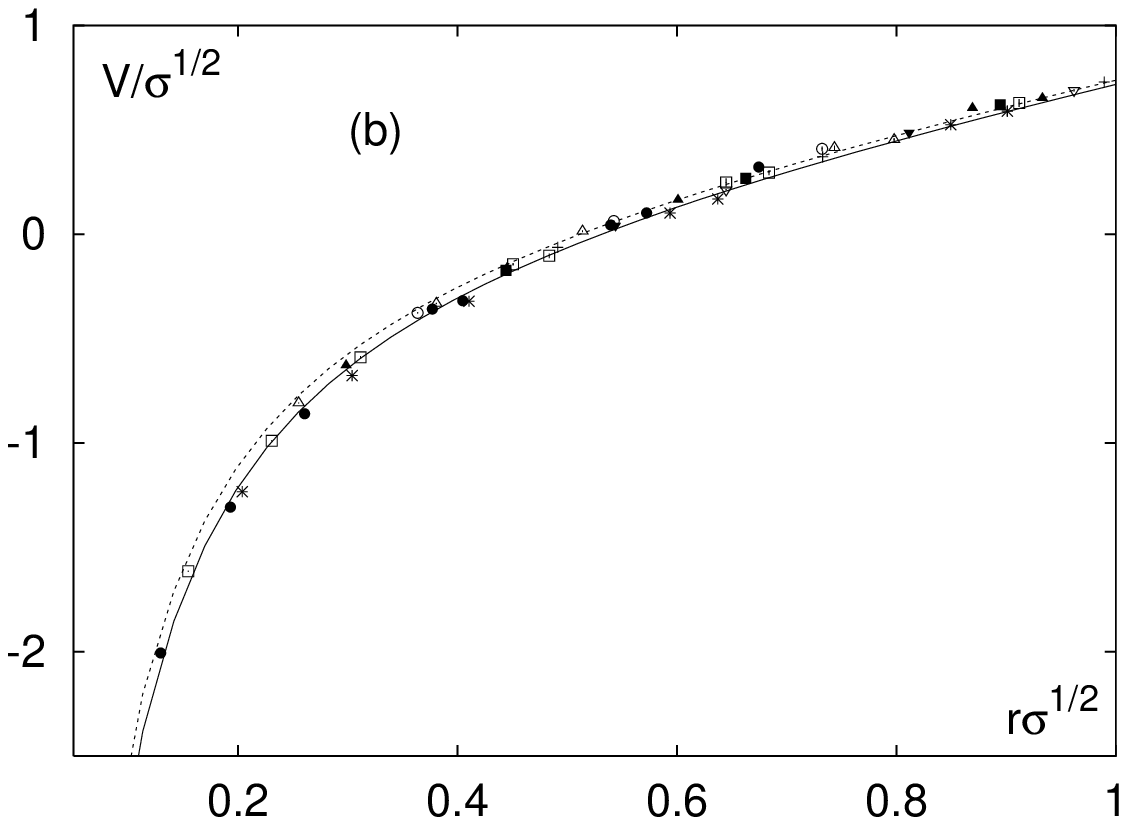,width=8.65cm}
\caption{
  (a) The heavy quark potential at $T=0$ from \cite{Karsch:2000kv} obtained
  from 2-flavor QCD lattice simulations with quark masses $ma=0.1$ for
  different values of the lattice coupling $\beta$. Fig.~2(b) shows an
  enlargement of the short distance distance regime. The data are matched to
  the bosonic string potential (dashed line) at large distances. Included is
  also the fit to the Cornell form (solid line) given in Eq.~(\ref{t=0ansatz}).
  Note here that the heavy quark potential from quenched lattice QCD and the
  string model potential coincide already at $r\sqrt{\sigma}\;\gsim\;0.8$
  \cite{Necco:2001xg,Luscher:2002qv} ($r\;\gsim\;0.4$ fm).  }
\label{peik}
\end{figure}

\subsection{Heavy quark potential at $T=0$}
For the determination of the heavy quark potential at zero temperature, $V(r)$,
we have used the measurements of large smeared Wilson loops given in
\cite{Karsch:2000kv} for the same simulation parameters ($N_f$=2 and $ma=0.1$)
and action. To eliminate the divergent self-energy contributions we matched
these data for all $\beta$-values (different $\beta$-values correspond to
different values of the lattice spacing $a$) at large distances to the bosonic
string potential,
\begin{eqnarray}
V(r) &=& - \frac{\pi}{12}\frac{1}{r} + \sigma r \nonumber\\
&\equiv&-\frac{4}{3}\frac{\alpha_{\text{str}}}{r}+\sigma r\;,
\label{string-cornell}
\end{eqnarray}
where we already have separated the Casimir factor so that
$\alpha_{\text{str}}\equiv\pi/16$. In this normalization any divergent
contributions to the lattice potential are eliminated uniquely. In
Fig.~\ref{peik} we show our results together with the heavy quark potential
from the string picture (dashed line). One can see that the data are well
described by Eq.~(\ref{string-cornell}) at large distances, {\em i.e.}
$r\sqrt{\sigma}\;\gsim\;0.8$, corresponding to $r\;\gsim\;0.4$ fm. At these
distances we see no major difference between the 2-flavor QCD potential
obtained from Wilson loops and the quenched QCD potential which can be well
parameterized within the string model already for $r\;\gsim\;0.4$ fm
\cite{Necco:2001xg,Luscher:2002qv}. In fact, we also do not see any signal for
string breaking in the zero temperature QCD heavy quark potential. This is
expected due to the fact that the Wilson loop operator used here for the
calculation of the $T=0$ potential has only small overlap with states where
string breaking occurs \cite{Bernard:2001tz,Pennanen:2000yk}. Moreover, the
distances for which we analyze the data for the QCD potential are below
$r\;\lsim\;1.2$ fm at which string breaking is expected to set in at zero
temperature and similar quark masses \cite{Pennanen:2000yk}.

\subsection{The coupling at $T=0$}\label{couplt=0}
Deviations from the string model and from the pure gauge potential, however,
are clearly expected to become apparent in the 2-flavor QCD potential at small
distances and may already be seen from the short distance part in
Fig.~\ref{peik}. These deviations are expected to arise from an asymptotic
weakening of the QCD coupling, {\em i.e.} $\alpha=\alpha(r)$, and to some
extent also due to the effect of including dynamical quarks, {\em i.e.} from
leading order perturbation theory one expects
\begin{eqnarray}
\alpha(r) \simeq \frac{1}{8\pi} \frac{1}{\beta_0 \log \left(1/(r \Lambda_{\text{ QCD}})\right)}\;,
\label{runningcoupling}
\end{eqnarray}
with
\begin{eqnarray}
\beta_0 = \frac{33-2N_f}{48 \pi^2}\;,
\end{eqnarray}
where $N_f$ is the number of flavors and $\Lambda_{\text{QCD}}$ denotes the
corresponding QCD-$\Lambda$-scale. The data in Fig.~\ref{peik}(b) show a slightly
steeper slope at distances below $r\sqrt{\sigma}\simeq0.5$ compared to the pure
gauge potential given in Ref.~\cite{Necco:2001xg} indicating that the QCD
coupling gets stronger in the entire distance range analyzed here when
including dynamical quarks. This is in qualitative agreement with
(\ref{runningcoupling}). To include the effect of a stronger Coulombic part in
the QCD potential we test the Cornell parameterization,
\begin{eqnarray}
\frac{V(r)}{\sqrt{\sigma}} = -\frac{4}{3}\frac{\alpha}{r\sqrt{\sigma}} + r \sqrt{\sigma}
\label{t=0ansatz}\;,
\end{eqnarray}
with a free parameter $\alpha$. From a best-fit analysis of Eq.~(\ref{t=0ansatz})
to the data ranging from $0.2\;\lsim\;r\sqrt{\sigma}\;\lsim\;2.6$ we find
\begin{eqnarray}
\alpha&=&0.212 (3)\;.\label{res}
\end{eqnarray}
This already may indicate that the logarithmic weakening of the coupling with
decreasing distance will not too strongly influence the properties of the QCD
potential at these distances, {\em i.e.} at $r\;\gsim\;0.1$ fm. However, the
value of $\alpha$ is moderately larger than
$\alpha_{\text{str}}\;\simeq\;0.196$ introduced above. To compare the relative
size of $\alpha$ in full QCD to $\alpha$ in the quenched theory we again have
performed a best-fit analysis of the quenched zero temperature potential given
in \cite{Necco:2001xg} using the Ansatz given in Eq.~(\ref{t=0ansatz}) and a
similar distance range. Here we find $\alpha_{\text{quenched}} = 0.195(1)$
which is again smaller than the value for the QCD coupling but quite comparable
to $\alpha_{\text{str}}$. In earlier studies of the heavy quark potentials in
pure gauge theories and full QCD even larger values for the couplings were
reported
\cite{Glassner:1996xi,Allton:1998gi,Aoki:1998sb,Bali:2000vr,AliKhan:2001tx,Aoki:2002uc}.
To avoid here any confusions concerning the value of $\alpha$ we should stress
that $\alpha$ should not be mixed with some value for the QCD coupling constant
$\alpha_{QCD}$, it simply is a fit parameter indicating the 'average strength'
of the Coulomb part in the Cornell potential.  The QCD coupling could be
identified properly only in the entire perturbative distance regime and will be
a running coupling, {\em i.e.}  $\alpha_{\text{QCD}}=\alpha_{\text{QCD}}(r)$.

When approaching the short distance perturbative regime the Cornell form will
overestimate the value of the coupling due to the perturbative logarithmic
weakening of the latter, $\alpha_{\text{QCD}}=\alpha_{\text{QCD}}(r)$.  To
analyze the short distance properties of the QCD potential and the coupling in
more detail, {\em i.e.} for $r\;\lsim\;0.4$ fm, and to firmly establish here
the onset of its perturbative weakening with decreasing distance, it is
customary to do so using non-perturbative definitions of running couplings.
Following the discussions on the running of the QCD coupling
\cite{Bali:1992ru,Peter:1997me,Schroder:1998vy,Necco:2001xg,Necco:2001gh}, it
appears most convenient to study the QCD force, {\em i.e.} $dV(r)/dr$, rather
than the QCD potential. In this case one defines the QCD coupling in the
so-called $qq$-scheme,
\begin{eqnarray}
\alpha_{qq}(r)&\equiv&\frac{3}{4}r^2\frac{dV(r)}{dr}\;.
\label{alp_qq}
\end{eqnarray} 
In this scheme any undetermined constant contribution to the heavy quark
potential cancels out. Moreover, the large distance, non-perturbative
confinement contribution to $\alpha_{qq}(r)$ is positive and allows for a
smooth matching of the perturbative short distance coupling to the
non-perturbative large distance confinement signal. In any case, however, in
the non-perturbative regime the value of the coupling will depend on the
observable used for its definition.

\begin{figure}[tbp]
  \epsfig{file=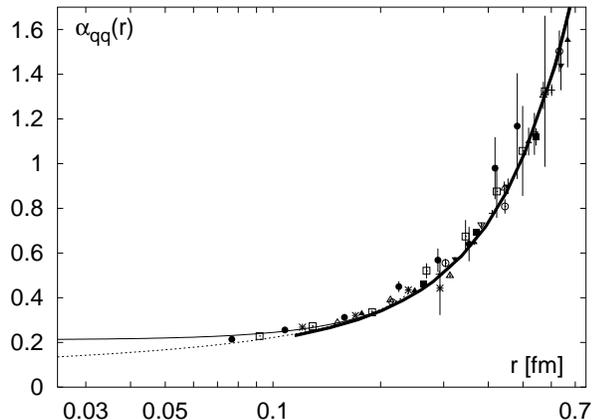,width=8.65cm}
\caption{
  The short distance part of the running coupling $\alpha_{qq}(r)$ in 2-flavor
  QCD at zero temperature defined in Eq.~(\ref{alp_qq}) as function of the
  distance $r$ (in physical units). The symbols for the different
  $\beta$-values are chosen as indicated in Fig.~\ref{peik}(a). The lines are
  discussed in the text.  }
\label{peiks}
\end{figure}
We have calculated the derivatives of the potential with respect to the
distance, $dV(r)/dr$, by using finite difference approximations for neighboring
distances on the lattice for each $\beta$-value separately.  Our results for
$\alpha_{qq}(r)$ as a function of distance in physical units for 2-flavor QCD
are summarized in Fig.~\ref{peiks}. The symbols for the $\beta$-values are
chosen as in Fig.~\ref{peik}(a). We again show in that figure the corresponding
line for the Cornell fit (solid line). At large distances, $r\;\gsim\;0.4$ fm,
the data clearly mimic the non-perturbative confinement part of the QCD force,
$\alpha_{qq}(r)\simeq3r^2\sigma/4$. We also compare our data to the recent high
statistics calculation in pure gauge theory (thick solid line)
\cite{Necco:2001xg}. These data are available for $r\;\gsim\;0.1$ fm and within
the statistics of the QCD data no significant differences could be identified
between the QCD and pure gauge data for $r\;\gsim\;0.4$ fm. At smaller
distances ($r\;\lsim\;0.4$ fm), however, the data show some enhancement
compared to the coupling in quenched QCD. The data below $0.1$ fm, moreover,
fall below the large distance Cornell fit. This may indicate the logarithmic
weakening of the coupling. At smaller distances than $0.1$ fm we therefore
expect the QCD potential to be influenced by the weakening of the coupling and
$\alpha_{qq}(r)$ will approach values clearly smaller than $\alpha$ deduced
from the Cornell Ansatz. Unfortunately we can, at present, not go to smaller
distances to clearly demonstrate this behavior with our data in 2-flavor QCD.
Moreover, at small distances cut-off effects may also influence our analysis of
the coupling and more detailed studies are required here. Despite these
uncertainties, however, in earlier studies of the coupling in pure gauge theory
\cite{Necco:2001xg,Necco:2001gh,Kaczmarek:2004gv} it is shown that the
perturbative logarithmic weakening becomes already important at distances
smaller than $0.2$ fm and contact with perturbation theory could be
established.

As most of our lattice data for the finite temperature quark anti-quark free
energies do not reach distances smaller than $0.1$ fm we use in the following
the Cornell form deduced in (\ref{t=0ansatz}) as reference to the zero
temperature heavy quark potential.
   
\section{Quark anti-quark free energy}\label{secfreee}
We will analyze here the temperature dependence of the change in free energy
due to the presence of a heavy (static) quark anti-quark pair in a 2-flavor QCD
heat bath.  The static quark sources are described by the Polyakov loop,
\begin{eqnarray}
L(\vec{x})&=&\frac{1}{3}\tr W(\vec{x})\;,\label{pol}
\end{eqnarray}
with
\begin{eqnarray}
W(\vec{x}) = \prod_{\tau=1}^{N_\tau} U_0(\vec{x},\tau)\;,\label{loop}
\end{eqnarray} 
where we already have used the lattice formulation with $U_0(\vec{x},\tau) \in
SU(3)$ being defined on the lattice link in time direction. The change in free
energy due to the presence of the static color sources in color singlet ($F_1$)
and color octet ($F_8$) states can be calculated in terms of Polyakov loop
correlation functions
\cite{McLerran:1981pb,Philipsen:2002az,Nadkarni:1986as,Nadkarni:1986cz},
\begin{eqnarray}
e^{-F_1(r)/T+C}&=&\frac{1}{3} \tr \langle  W(\vec{x}) W^{\dagger}(\vec{y}) \rangle 
\label{f1}\;,\\
e^{-F_8(r)/T+C}&=&\frac{1}{8}\langle \tr  W(\vec{x}) \tr W^{\dagger}(\vec{y})\rangle- \nonumber \\
&&
                \frac{1}{24} \tr \langle W(\vec{x}) W^{\dagger}(\vec{y}) \rangle\; ,
\label{f8}
\end{eqnarray}
where $r=|\vec{x}-\vec{y}|$. As it stands, the correlation functions for the
color singlet and octet free energies are gauge dependent quantities and thus
gauge fixing is needed to define them properly. Here, we follow
\cite{Philipsen:2002az} and fix to Coulomb gauge. In parts we also consider the
so-called color averaged free energy defined through the manifestly gauge
independent correlation function of two Polyakov loops,
\begin{eqnarray}
e^{-F_{\bar q q}(r)/T+C}&=&\frac{1}{9}\langle \tr W(\vec{x}) \tr W^{\dagger}(0) \rangle \nonumber\\
&=&\langle L(\vec{x})L^\dagger(\vec{y})\rangle\; .
\label{fav}
\end{eqnarray}

The constant $C$ appearing in (\ref{f1}), (\ref{f8}) and (\ref{fav}) also
includes divergent self-energy contributions which require renormalization.
Following \cite{Kaczmarek:2002mc} the free energies have been normalized such
that the color singlet free energy approaches the heavy quark potential (solid
line) at the smallest distance available on the lattice, $F_1(r/a=1, T)=V(r)$.
In Sec.~\ref{renormalization} we will explain the connection of this procedure
to the the renormalized Polyakov loop and show the resulting renormalization
constants in Table~\ref{tab:ren}.

\begin{figure}[tbp]
  \epsfig{file=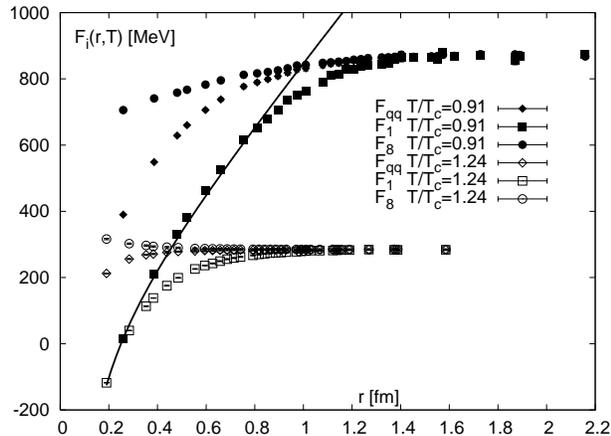,width=8.65cm}
\caption{
  Heavy quark free energies for $2$ flavors of dynamical quarks at a quark mass
  of $m/T=0.4$ calculated on $16^3\times 4$ lattices. Shown are the free
  energies in different color channels, the singlet ($F_1$), octet ($F_8$) and
  color averaged ($F_{\bar q q}$) free energies normalized here as discussed in
  \cite{Kaczmarek:2002mc} to the zero temperature potential obtained in
  Sec.~\ref{sect0} (solid line).  }
\label{saos}
\end{figure}
Some results for the color singlet, octet and averaged quark anti-quark free
energies are shown in Fig.~\ref{saos} for one temperature below and one
temperature above deconfinement, respectively.  The free energies calculated in
different color channels coincide at large distances and clearly show the
effects from string breaking below and color screening above deconfinement. The
octet free energies above $T_c$ are repulsive for all distances while below
$T_c$ the distances analyzed here are not small enough to show the
(perturbatively) expected repulsive short distance part. Similar results are
obtained at all temperatures analyzed here. In the remainder of this section we
study in detail the thermal modifications of these free energies from short to
large distances. We begin our analysis of the free energies at small distances
in Sec.~\ref{couplatt} with a discussion of the running coupling which leads to
the renormalization of the free energies in Sec.~\ref{renormalization}. The
separation of small and large distances which characterizes sudden qualitative
changes in the free energy will be discussed in Sec.~\ref{secshort}. Large
distance modifications of the quark anti-quark free energy will be studied in
Sec.~\ref{colorscreening} at temperatures above and in
Sec.~\ref{stringbreaking} at temperatures below deconfinement.
 
Our analysis of thermal modifications of the strong interaction will mainly be
performed for the color singlet free energy. In this case a rather simple
Coulombic $r$-dependence is suggested by perturbation theory at $T=0$ and short
distances as well as for large distances at high temperatures. In particular, a
proper $r$-dependence of $F_{\bar q q}$ is difficult to establish
\cite{Kaczmarek:2002mc}. This maybe is attributed to contributions from higher
excited states \cite{Jahn:2004qr} or to the repulsive contributions from states
with static charges fixed in an octet configuration.

\subsection{The running coupling at $T\neq0$}\label{couplatt}
\begin{figure}[tbp]
  \epsfig{file=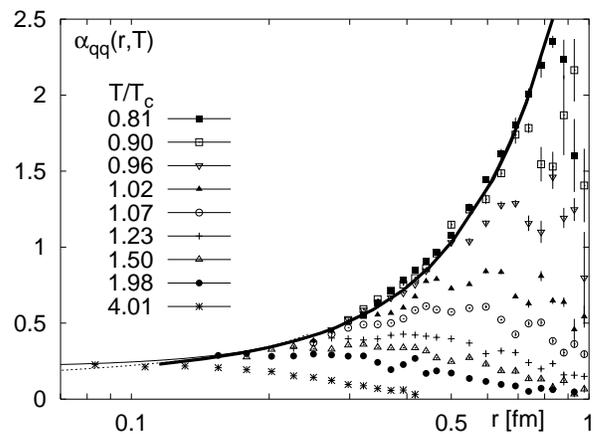,width=8.65cm}
\caption{
  The running coupling in the $qq$-scheme defined in Eq.~(\ref{alp_rT})
  calculated from derivatives of the color singlet free energies with respect
  to $r$ at several temperatures as function of distance below and above
  deconfinement. We also show the corresponding coupling at zero temperature
  (solid line) from Eq.~(\ref{t=0ansatz}) and compare the results again to the
  results in pure gauge theory (thick solid and dashed lines)
  \cite{Necco:2001gh,Necco:2001xg}.  }
\label{couplt}
\end{figure}
We extend here our studies of the coupling at zero temperature to finite
temperatures below and above deconfinement following the conceptual approach
given in \cite{Kaczmarek:2004gv}. In this case the appropriate observable is
the color singlet quark anti-quark free energy and its derivative. We use the
perturbative short and large distance relation from one gluon exchange
\cite{Nadkarni:1986as,Nadkarni:1986cz,McLerran:1981pb}, {\em i.e.} in the limit
$r\Lambda_{\text{QCD}}\ll1$ zero temperature perturbation theory suggests
\begin{eqnarray}
F_1(r,T)\;\equiv\;V(r)&\simeq&-\frac{4}{3}\frac{\alpha(r)}{r}\;,\label{alp_rT1}
\end{eqnarray}
while high temperature perturbation theory, {\em i.e.} $rT\gg1$ and $T$ well
above $T_c$, yields
\begin{eqnarray}
F_1(r,T)&\simeq&-\frac{4}{3}\frac{\alpha(T)}{r}e^{-m_D(T)r}\;.\label{alp_rT2}
\end{eqnarray}
In both relations we have neglected any constant contributions to the free
energies which, in particular, at high temperatures will dominate the large
distance behavior of the free energies. Moreover, we already anticipated here
the running of the couplings with the expected dominant scales $r$ and $T$ in
both limits. At finite temperature we define the running coupling in analogy to
$T=0$ as (see \cite{Kaczmarek:2002mc,Kaczmarek:2004gv}),\begin{eqnarray}
  \alpha_{qq}(r,T)&\equiv&\frac{3}{4}r^2 \frac{dF_1(r,T)}{dr}\;.\label{alp_rT}
\end{eqnarray}  
With this definition any undetermined constant contributions to the free
energies are eliminated and the coupling defined here at finite temperature
will recover the coupling at zero temperature defined in (\ref{alp_qq}) in the
limit of small distances. Therefore $\alpha_{qq}(r,T)$ will show the (zero
temperature) weakening in the short distance perturbative regime. In the large
distance limit, however, the coupling will be dominated by Eq.~(\ref{alp_rT2})
and will be suppressed by color screening,
$\alpha_{qq}(r,T)\simeq\alpha(T)\exp(-m_D(T)r)$, $rT\gg1$. It thus will exhibit
a maximum at some intermediate distance. Although in the large distance regime
$\alpha_{qq}(r,T)$ will be suppressed by color screening and thus
non-perturbative effects will strongly control the value of $\alpha_{qq}(r,T)$,
in this limit the temperature dependence of the coupling, $\alpha(T)$, can be
extracted by directly comparing the singlet free energy with the high
temperature perturbative relation above deconfinement. Results from such an
analysis will be given in Sec.~\ref{colorscreening}.

We calculated the derivative, $dF_1/dr$, of the color singlet free energies
with respect to distance by using cubic spline approximations of the
$r$-dependence of the free energies for each temperature. We then performed the
derivatives on basis of these splines. Our results for $\alpha_{qq}(r,T)$
calculated in this way are shown in Fig.~\ref{couplt} and are compared to the
coupling at zero temperature discussed already in Sec.~\ref{couplt=0}. Here the
thin solid line corresponds to the coupling in the Cornell Ansatz deduced in
Eq.~(\ref{t=0ansatz}). We again show in this figure the results from
$SU(3)$-lattice (thick line) and perturbative (dashed line) calculations at
zero temperature from \cite{Necco:2001gh,Necco:2001xg}. The strong
$r$-dependence of the running coupling near $T_c$ observed already in pure
gauge theory \cite{Kaczmarek:2004gv} is also visible in 2-flavor QCD.  Although
our data for 2-flavor QCD do not allow for a detailed quantitative analysis of
the running coupling at smaller distances, the qualitative behavior is in quite
good agreement with the recent quenched results. At large distances the running
coupling shows a strong temperature dependence which sets in at shorter
separations with increasing temperature. At temperatures close but above $T_c$,
$\alpha_{qq}(r,T)$ coincides with $\alpha_{qq}(r)$ already at separations
$r\;\simeq\;0.4$ fm and clearly mimics here the confinement part of
$\alpha_{qq}(r)$. This is also apparent in quenched QCD
\cite{Kaczmarek:2004gv}.  Remnants of the confinement part of the QCD force may
survive the deconfinement transition and could play an important role for the
discussion of non-perturbative aspects of quark anti-quark interactions at
temperatures moderately above $T_c$ \cite{Shuryak:2004tx,Brown:2004qi}. A clear
separation of the different effects usually described by the concept of color
screening ($T\;\gsim\;T_c$) and effects usually described by the concept of
string-breaking ($T\;\lsim\;T_c$) is difficult to establish at temperatures in
the close vicinity of the confinement deconfinement crossover.

\begin{figure}[tbp]
  \epsfig{file=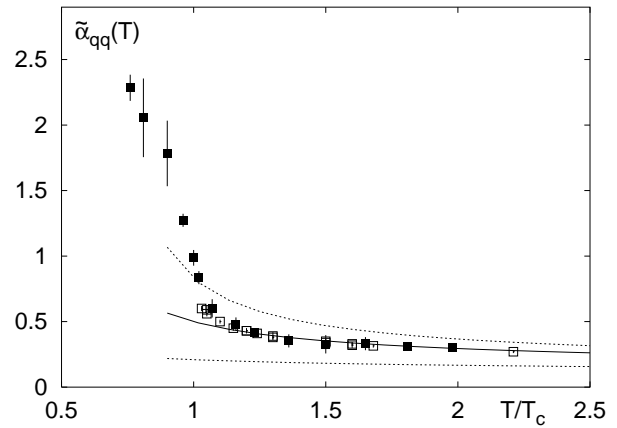,width=8.65cm}
\caption{
  The size of the maximum, $\tilde{\alpha}_{qq}(T)$, defined in
  Eq.~(\ref{alp_Tdef}), as function of temperature in $2$-flavor QCD (filled
  symbols) and pure gauge theory (open symbols) from \cite{Kaczmarek:2004gv}.
  The lines are explained in the text.  }
\label{alp_qqT}
\end{figure}

We also analyzed the size of the maximum that the running coupling
$\alpha_{qq}(r,T)$ at fixed temperature exhibits at a certain distance,
$r_{max}$, {\em i.e.} we identify a temperature dependent coupling,
$\tilde{\alpha}_{qq}(T)$, defined as
\begin{eqnarray}
\tilde{\alpha}_{qq}(T)&\equiv&\alpha_{qq}(r_{max},T)\;.\label{alp_Tdef} 
\end{eqnarray}
The values for $r_{max}$ will be discussed in Sec~\ref{secshort} (see
Fig.~\ref{onset}). Values for $\tilde{\alpha}_{qq}(T)$ are also available in
pure gauge theory \cite{Kaczmarek:2004gv} at temperatures above
deconfinement\footnote{In pure gauge theory $r_{max}$ and
  $\tilde{\alpha}_{qq}(T)$ would be infinite below $T_c$.}.  
Our results for $\tilde{\alpha}_{qq}(T)$ in
$2$-flavor QCD and pure gauge theory are shown in Fig.~\ref{alp_qqT} as
function of temperature, $T/T_c$. At temperatures above deconfinement we cannot
identify significant differences between the data from pure gauge and 2-flavor
QCD\footnote{Note here, however, the change in temperature scale from $T_c=202$
 MeV in full and $T_c=270$ MeV in quenched QCD.}. 
Only at temperatures quite close but above the phase
transition small differences between full and quenched QCD become visible in
$\tilde{\alpha}_{qq}(T)$. Nonetheless, the value of $\tilde{\alpha}_{qq}(T)$
drops from about $0.5$ at temperatures only moderately larger than the
transition temperature, $T\;\gsim\;1.2T_c$, to a value of about $0.3$ at
$2T_c$. This change in $\tilde{\alpha}_{qq}(T)$ with temperature calculated in
$2$-flavor QCD does not appear to be too dramatic and might indeed be described
by the $2$-loop perturbative coupling,
\begin{eqnarray}
g_{\text{2-loop}}^{-2}(T)=2\beta_0\ln\left(\frac{\mu T}
{\Lambda_{\overline{MS}}}\right)+\frac{\beta_1}{\beta_0}
\ln\left(2\ln\left(\frac{\mu T}{\Lambda_{\overline{MS}}}\right)\right),\nonumber\\
\label{2loop}
\end{eqnarray}  
with
\begin{eqnarray}
\beta_0&=&\frac{1}{16\pi^2}\left(11-\frac{2N_f}{3}\right)\;,\nonumber\\
\beta_1&=&\frac{1}{(16\pi^2)^2}\left(102-\frac{38N_f}{3}\right)\;,\nonumber
\end{eqnarray}
assuming vanishing quark masses. In view of the ambiguity in setting the scale
in perturbation theory, $\mu T$, we performed a best-fit analysis to fix the
scale for the entire temperature range, $1.2\;\lsim\;T/T_c\;\lsim\;2$. We find
here $\mu=1.14(2)\pi$ with $T_c/\Lambda_{\overline{MS}}=0.77(21)$ using
$T_c\simeq202(4)$ MeV \cite{Karsch:2000ps} and $\Lambda_{\overline{MS}}\simeq
261(17)$ MeV \cite{Gockeler:2005rv}, which is still in agreement with the lower
limit of the range of scales one commonly uses to fix perturbative couplings,
$\mu=\pi,...,4\pi$. This is shown by the solid line (fit) in Fig.~\ref{alp_qqT}
including the error band estimated through $\mu=\pi$ to $\mu=4\pi$ and the
error on $T_c/\Lambda_{\overline{MS}}$ (dotted lines). We will turn back to a
discussion of the temperature dependence of the coupling above deconfinement in
Sec.~\ref{colorscreening}.

At temperatures in the vicinity and below the phase transition temperature,
$T\;\lsim\;1.2T_c$, the behavior of $\tilde{\alpha}_{qq}(T)$ is, however, quite
different from the perturbative logarithmic change with temperature. The values
for $\tilde{\alpha}_{qq}(T)$ rapidly grow here with decreasing temperature and
approach non-perturbatively large values. This again shows that
$\tilde{\alpha}_{qq}(r,T)$ mimics the confinement part of the zero temperature
force still at relatively large distances and that this behavior persists up to
temperatures close but above deconfinement. This again demonstrates the
persistence of confinement forces at $T\;\gsim\;T_c$ and intermediate distances
and demonstrates the difficulty to separate clearly the different effects
usually described by color screening and string breaking in the vicinity of the
phase transition. We note here, however, that similar to the coupling in
quenched QCD \cite{Kaczmarek:2004gv} the coupling which describes the short
distance Coulombic part in the free energies is almost temperature independent
in this temperature regime, {\em i.e.} even at relatively large distances the
free energies shown in Fig.~\ref{fes} show no or only little temperature
dependence below deconfinement.

\subsection{Renormalization of the quark anti-quark free energies and Polyakov
  loop}\label{renormalization} On the lattice the expectation value of the
Polyakov loop and its correlation functions suffer from linear divergences.
This leads to vanishing expectation values in the continuum limit, $a\to0$, at
all temperatures. To become a meaningful physical observable a proper
renormalization is required
\cite{Kaczmarek:2002mc,Dumitru:2004gd,deForcrand:2001nd}.  We follow here the
conceptual approach suggested in \cite{Kaczmarek:2002mc,Phd} and extend our
earlier studies in pure gauge theory to the present case of 2-flavor QCD. First
experiences with this renormalization method in full QCD were already reported
in \cite{Kaczmarek:2003ph,Petreczky:2004pz}.

In the limit of short distances, $r\ll1/T$, thermal modifications of the quark
anti-quark free energy become negligible and the running coupling is controlled
by distance only.  Thus we can fix the free energies at small distances to the
heavy quark potential, $F_1(r\ll1/T,T)\simeq V(r)$, and the renormalization
group equation (RGE) will lead to
\begin{eqnarray}
\lim_{r\to0}T\frac{dF_1(r,T)}{dT}&=&0\;,\label{RGEf}
\end{eqnarray}  
where we already have assumed that the continuum limit, $a\to0$, has been
taken. On basis of the analysis of the coupling in Sec.~\ref{couplatt} and
experiences with the quark anti-quark free energy in pure gauge theory
\cite{Kaczmarek:2002mc,Kaczmarek:2004gv} we assume here that the color singlet
free energies in 2-flavor QCD calculated on finite lattices with temporal
extent $N_\tau=4$ already have approached appropriate small distances,
$r\ll1/T$, allowing for renormalization.

The (renormalized) color singlet quark anti-quark free energies, $F_1(r,T)$,
and the heavy quark potential, $V(r)$ (line), were already shown in
Fig.~\ref{fes}(a) as function of distance at several temperatures close to the
phase transition. From that figure it can be seen that the quark anti-quark
free energy fixed at small distances approaches finite, temperature dependent
plateau values at large distances signaling color screening ($T\;\gsim\;T_c$)
and string breaking ($T<T_c$). These plateau values, $F_\infty(T)\equiv
F_1(r\to\infty,T)$, are decreasing with increasing temperature in the
temperature range analyzed here. In general it is expected that $F_\infty(T)$
will continue to increase at smaller temperature and will smoothly match
$V(r\equiv\infty)$ \cite{Digal:2001iu} at zero temperature while it will become
negative at high temperature and asymptotically is expected to become
proportional to $g^3T$ \cite{Gava:1981qd,Kaczmarek:2002mc}. The plateau value
of the quark anti-quark free energy at large distances can be used to define
non-perturbatively the renormalized Polyakov loop \cite{Kaczmarek:2002mc}, {\em
  i.e.}
\begin{eqnarray}
L^{\text{ren}}(T)&=&\exp\left(-\frac{F_\infty(T)}{2T}\right)\;.\label{renPloop}
\end{eqnarray}      
As the unrenormalized free energies approach $|\langle L\rangle|^2$ at large
distances, this may be reinterpreted in terms of a renormalization constant
that has been determined by demanding (\ref{RGEf}) to hold at short distances
\cite{Dumitru:2003hp,Zantow:2003uh},
\begin{eqnarray}
L^{\text{ren}}&\equiv&|\langle \left(Z(g,m)\right)^{N_\tau} L \rangle|\;.
\end{eqnarray} 
The values for $Z(g,m)$ for our simulation parameters are summarized in
Table~\ref{tab:ren}. The normalization constants for the free energies
appearing in (\ref{f1}-\ref{fav}) are then given by
\begin{eqnarray}
C=-2 N_\tau Z(g,m).
\end{eqnarray}
An analysis of the renormalized Polyakov loop expectation value in high
temperature perturbation theory \cite{Gava:1981qd} suggests at (resummed)
leading order\footnote{In Ref.~\cite{Gava:1981qd} the Polyakov loop expectation
  value is calculated in pure gauge theory and the Debye mass,
  $m_D(T)/T=\sqrt{N_c/3}g(T)$, enters here through the resummation of the
  gluon polarization tensor. When changing from pure gauge to full QCD quark
  loops will contribute to the polarization tensor. In this case resummation
  will lead to the Debye mass given in (\ref{LOscreen}). Thus the flavor
  dependence in Eq.~(\ref{lrenpert}) at this level is given only by the Debye
  mass.}, the behavior
\begin{eqnarray}
L^{\text{ren}}(T)&\simeq&1+\frac{2}{3}\frac{m_D(T)}{T}\alpha(T)\label{lrenpert}\;
\end{eqnarray} 
in the fundamental representation. Thus high temperature perturbation theory
suggests that the limiting value at infinite temperature,
$L^{\text{ren}}(T\to\infty)=1$ is approached from above. An expansion of
(\ref{renPloop}) then suggests
$F_\infty(T)\simeq-\frac{4}{3}m_D(T)\alpha(T)\simeq-{\cal O}(g^3T)$. We thus
expect $F_\infty(T)\to-\infty$ in the high temperature limit.

\begin{figure}[tbp]
  \epsfig{file=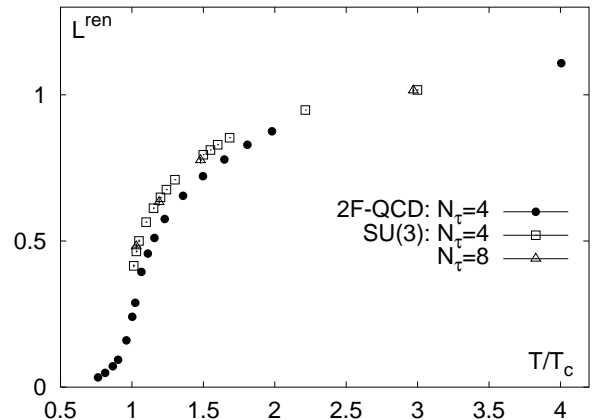,width=8.65cm}
\caption{
  The renormalized Polyakov loop in $2$-flavor QCD (filled circles) compared to
  the quenched results (open symbols) from \cite{Kaczmarek:2002mc} as function
  of temperature in units of the (different) transition temperatures.  }
\label{renpol}
\end{figure}

\begin{table}[tbp]
\centering
\setlength{\tabcolsep}{0.7pc}
\begin{tabular}{|ll|ll|}
\hline
$\beta$ & $Z(g,m)$ & $T/T_c$ & $L^{\text{ren}}(T)$ \\
\hline
3.52   & 1.333(19)  &   0.76  &   0.033(2)     \\
3.55   & 1.351(10)  &   0.81  &   0.049(2)     \\
3.60   & 1.370(08)  &   0.90  &   0.093(2)     \\
3.63   & 1.376(07)  &   0.96  &   0.160(3)     \\
3.65   & 1.376(07)  &   1.00  &   0.241(5)     \\
3.66   & 1.375(06)  &   1.02  &   0.290(5)     \\
3.68   & 1.370(06)  &   1.07  &   0.398(7)     \\
3.72   & 1.374(02)  &   1.16  &   0.514(3)     \\
3.75   & 1.379(02)  &   1.23  &   0.575(2)     \\
3.80   & 1.386(01)  &   1.36  &   0.656(2)     \\
3.85   & 1.390(01)  &   1.50  &   0.722(2)     \\
3.90   & 1.394(01)  &   1.65  &   0.779(1)     \\
3.95   & 1.396(13)  &   1.81  &   0.828(3)     \\
4.00   & 1.397(01)  &   1.98  &   0.874(1)     \\
4.43   & 1.378(01)  &   4.01  &   1.108(2)     \\
\hline
\end{tabular}
\caption{Renormalization constants, $Z(g,m)$, versus $\beta$ and the
  renormalized Polyakov loop, $L^{\text{ren}}$, versus $T/T_c$ for 2-flavor QCD
  with quark mass $m/T=0.4$.}
\smallskip
\label{tab:ren}
\end{table}
To avoid here any fit to the complicated $r$- and $T$-dependence of the quark
anti-quark free energy we estimate the value of $F_\infty(T)$ from the quark
anti-quark free energies at the largest separation available on a finite
lattice, $r=N_\sigma/2$. As the free energies in this renormalization scheme
coincide at large distances in the different color channels we determine
$F_\infty(T)$ from the color averaged free energies, {\em i.e.}
$F_\infty(T)\equiv F_{\bar q q}(r=N_\sigma/2,T)$. This is a manifestly gauge
invariant quantity.  In Fig.~\ref{renpol} we show the results for
$L^{\text{ren}}$ in 2-flavor QCD (filled symbols) compared to the quenched
results (open symbols) obtained in \cite{Kaczmarek:2002mc}.  In quenched QCD
$L^{\text{ren}}$ is zero below $T_c$ as the quark anti-quark free energy
signals permanent confinement, {\em i.e.}  $F_\infty(T\;\lsim\; T_c)=\infty$ in
the infinite volume limit, while it jumps to a finite value just above $T_c$.
The singularity in the temperature dependence of $L^{\text{ren}}(T)$ located at
$T_c$ clearly signals the first order phase transition in $SU(3)$ gauge theory.
The renormalized Polyakov loop in $2$-flavor QCD, however, is no longer zero
below $T_c$. Due to string breaking the quark anti-quark free energies approach
constant values leading to non-zero values of $L^{\text{ren}}$. Although the
renormalized Polyakov loop calculated in full QCD is no longer an order
parameter for the confinement deconfinement phase transition, it still shows a
quite different behavior in the two phases and a clear signal for a qualitative
change in the vicinity of the transition. Above deconfinement
$L^{\text{ren}}(T)$ yields finite values also in quenched QCD.

In the temperature range $1\;\lsim\;T/T_c\;\lsim\;2$ we find that in 2-flavor
QCD $L^{\text{ren}}$ lies below the results in quenched QCD. This, however, may
change at higher temperatures. The value for $L^{\text{ren}}$ at $4T_c$ is
larger than unity and we find indication for
$L^{\text{ren}}_{\mathrm{2-flavor}}(4T_c)\;\gsim\;L^{\text{ren}}_{\mathrm{quenched}}(4T_c)$.
The properties of $L^{\text{ren}}$, however, clearly depend on the relative
normalization of the quark anti-quark free energies in quenched and full QCD.

\subsection{Short vs. large distances}\label{secshort}
Having discussed the quark anti-quark free energies at quite small distances
where no or only little temperature effects influence the behavior of the free
energies and at quite large distances where aside from $T$ no other scale
controls the free energy, we now turn to a discussion of medium effects at
intermediate distances. The aim is to gain insight into distance scales that
can be used to quantify at which distances temperature effects in the quark
anti-quark free energies set in and may influence the in-medium properties of
heavy quark bound states in the quark gluon plasma.
    
It can be seen from Fig.~\ref{fes}(a) that the color singlet free energy changes
rapidly from the Coulomb-like short distance behavior to an almost constant
value at large distances. This change reflects the in-medium properties of the
heavy quark anti-quark pair, {\em i.e.} the string-breaking property and color
screening. To characterize this rapid onset of in-medium modifications in the
free energies we introduced in Ref.~\cite{Kaczmarek:2002mc} a scale, $r_{med}$,
defined as the distance at which the value of the $T=0$ potential reaches the
value $F_\infty(T)$, {\em i.e.}
\begin{eqnarray}
V(r_{med})&\equiv&F_\infty(T)\;.\label{rmed}
\end{eqnarray}
As $F_\infty(T)$ is a gauge invariant observable this relation provides a
non-perturbative, gauge invariant definition of the scale $r_{med}$. While in
pure gauge theory the color singlet free energies signal permanent confinement
at temperatures below $T_c$ leading to a proper definition of this scale only
above deconfinement, in full QCD it can be deduced in the whole temperature
range. On the other hand, the change in the coupling $\alpha_{qq}(r,T)$ as
function of distance at fixed temperature mimics the qualitative change in the
interaction when going from small to large distances and the coupling exhibits
a maximum at some intermediate distance. The location of this maximum,
$r_{max}$, can also be used to identify a scale that characterizes separation
between the short distance vacuumlike and the large distance medium modified
interaction between the static quarks \cite{Kaczmarek:2004gv}. Due to the rapid
crossover from short to large distance behavior (see Fig.~\ref{fes}(a)) it should
be obvious that $r_{med}$ and $r_{max}$ define similar scales, however, by
construction $r_{max}\;\lsim\;r_{med}$.

\begin{figure}[tbp]
  \epsfig{file=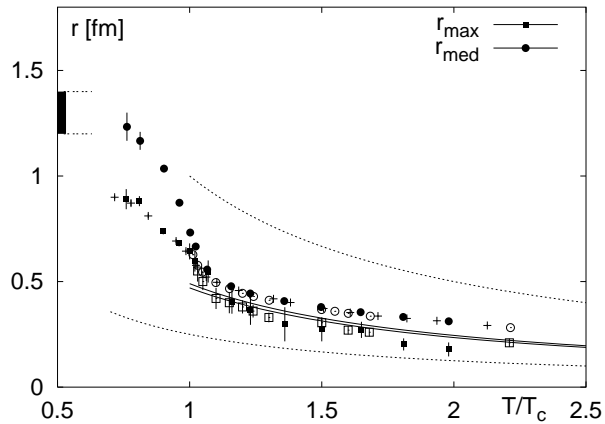,width=8.65cm}
\caption{
  The location of the maximum in $\alpha_{qq}(r,T)$ at fixed temperature,
  $r_{max}$ ($N_f$=0: open squares, $N_f$=2: filled squares), and our results
  for $r_{med}$ ($N_f$=0: open circles, $N_f$=2: filled circles, $N_f$=3:
  crosses) defined in Eq.~(\ref{rmed}) versus $T/T_c$. The band at the left frame
  indicates the distance range at which string breaking is expected to occur in
  $2$-flavor QCD at $T=0$ and quark mass $m_\pi/m_\rho\simeq0.7$
  \cite{Pennanen:2000yk}. The various lines are explained in the text.  }
\label{onset}
\end{figure}
To gain important information about the flavor and quark mass dependence of our
analysis of the scales in QCD, we also took data for $F_\infty(T)$ from
Ref.~\cite{Petreczky:2004pz} at smaller quark mass, $m_\pi/m_\rho\simeq0.4$
\cite{Peterpriv}, and calculated $r_{med}$ in $3$-flavor QCD with respect to
the parameterization of $V(r)$ given in \cite{Petreczky:2004pz}. It is
interesting to note here that a study of the flavor and quark mass dependence
of $r_{med}$ and $r_{max}$ is independent of any undetermined and maybe flavor
and/or quark mass dependent overall normalization of the corresponding $V(r)$
at zero temperature.  Our results for $r_{max}$ ($N_f$=0,2) and $r_{med}$
($N_f$=0,2,3) are summarized in Fig.~\ref{onset} as function of $T/T_c$. It can
be seen that the value $r_{max}\simeq 0.6$ fm is approached almost in common in
quenched and $2$-flavor QCD at the phase transition and it commonly drops to
about $0.25$ fm at temperatures about $2T_c$. No or only little differences
between $r_{max}$ calculated from pure gauge and $2$-flavor QCD could be
identified at temperatures above deconfinement. The temperature dependence of
$r_{med}$ is similar to that of $r_{max}$ and again we see no major differences
between pure gauge ($N_f$=0) and QCD ($N_f$=2,3) results. In the vicinity of
the transition temperature and above both scales almost coincide. In fact,
above deconfinement the flavor and finite quark mass dependence of $r_{med}$
appears quite negligible. At high temperature we expect $r_{med}\simeq1/gT$
\cite{Kaczmarek:2002mc} while in terms of $r_{max}$ we found agreement with
$r_{max}=0.48(1)$ fm $T_c/T$ (solid lines) at temperatures ranging up to
$12T_c$ \cite{Kaczmarek:2004gv}. Note that both scales clearly lie well above
the smallest distance attainable by us on the lattice, $rT\equiv1/N_\tau=1/4$.
This distance is shown by the lower dashed line in Fig.~\ref{onset}.

At temperatures below deconfinement $r_{max}$ and $r_{med}$ rapidly increase
and fall apart when going to smaller temperatures. In fact, at temperatures
below deconfinement we clearly see difference between $r_{med}$ calculated in
$2$- and $3$-flavor QCD. To some extend this is expected due to the smaller
quark mass used in the $3$-flavor QCD study as the string breaking energy gets
reduced. It is, however, difficult to clearly separate here a finite quark mass
effect from flavor dependence. In both cases $r_{med}$ approaches, already at
$T\simeq0.8T_c$, quite similar values to those reported for the distance where
string breaking at $T=0$ is expected at similar quark masses. In $2$-flavor QCD
at $T=0$ and quark mass $m_\pi/m_\rho\simeq0.7$ the string is expected to break
at about $1.2-1.4$ fm \cite{Pennanen:2000yk} while at smaller quark mass,
$m_\pi/m_\rho\simeq0.4$ it might break earlier \cite{Bernard:2001tz}.

\begin{figure}[tbp]
  \epsfig{file=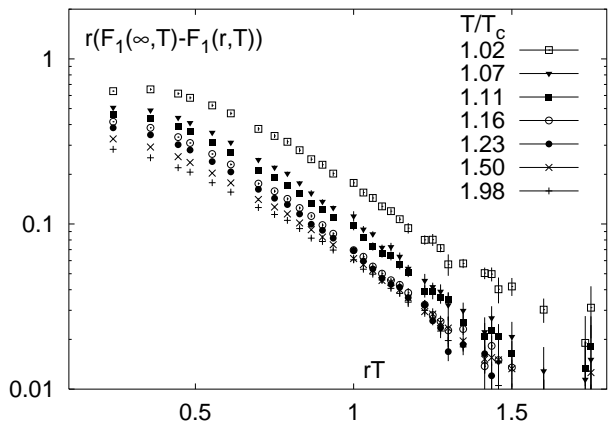,width=8.65cm}
\caption{
  The color singlet free energy versus $rT$ obtained from lattice calculations
  in $2$-flavor QCD at several temperatures above deconfinement. }
\label{screeningf}
\end{figure}

In contrast to the complicated $r$- and $T$-dependence of the free energy at
intermediate distances high temperature perturbation theory suggests a color
screened Coulomb behavior for the singlet free energy at large distances.  To
analyze this in more detail we show in Fig.~\ref{screeningf} the subtracted
free energies, $r(F_1(\infty,T)-F_1(r,T))$. It can be seen that this quantity
indeed decays exponentially at large distances, $rT\;\gsim\;1$. This allows us
to study the temperature dependence of the parameters $\alpha(T)$ and $m_D(T)$
given in Eq.~(\ref{alp_rT2}). At intermediate and small distances, however,
deviations from this behavior are expected and can clearly be seen and are to
some extent due to the onset of the $r$-dependence of the coupling at small
distances. These deviations from the simple exponential decay become important
already below some characteristic scale, $r_d$, which we can roughly identify
here as $r_dT\simeq0.8\;-\;1$. This scale which defines a lower limit for the
applicability of high temperature perturbation theory is shown by the upper
dashed line in Fig.~\ref{onset} ($r_dT=1$). It lies well above the scales
$r_{med}$ and $r_{max}$ which characterize the onset of medium modifications on
the quark anti-quark free energy.

\subsection{Screening properties above deconfinement and the coupling}\label{colorscreening}
\subsubsection{Screening properties and quarkonium binding}
We follow here the approach commonly used
\cite{Attig:1988ey,Nakamura:2004wr,Kaczmarek:2004gv} and define the
non-perturbative screening mass, $m_D(T)$, and the temperature dependent
coupling, $\alpha(T)$, from the exponential fall-off of the color singlet free
energies at large distances, $rT\;\gsim\;0.8\;$-$\;1$. A consistent definition
of screening masses, however, is accompanied by a proper definition of the
temperature dependent coupling and only at sufficiently high temperatures
contact with perturbation theory is expected
\cite{Kaczmarek:2004gv,Laine:2005ai}. A similar discussion of the color
averaged quark anti-quark free energy is given in
Refs.~\cite{Kaczmarek:1999mm,Petreczky:2001pd,Zantow:2001yf}.

\begin{figure}[tbp]
  \epsfig{file=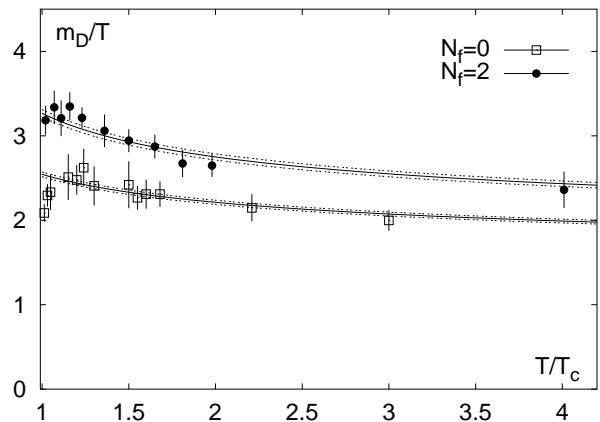,width=8.65cm}
\caption{
  The non-perturbatively defined screening masses from the large distance
  behavior of $F_1(r,T)$ calculated using $16^3\times4$ lattices as function of
  temperature in $2$-flavor QCD (filled circles) and in pure gauge theory (open
  squares \cite{Kaczmarek:2004gv}) using $32^3\times4$ lattices. The solid
  lines show the fit given in Eq.~(\ref{screenfit}) with the corresponding error
  band (dotted lines).  }
\label{screenmass}
\end{figure}

We used the Ansatz (\ref{alp_rT2}) to perform a best-fit analysis of the large
distance part of the color singlet free energies, {\em i.e.} we used fit
functions with the Ansatz
\begin{eqnarray}
F_1(r,T)-F_1(r=\infty,T) = - \frac{4a(T)}{3r}e^{-m(T)r},
\label{screenfit}
\end{eqnarray}
where the two parameters $a(T)$ and $m(T)$ are used to estimate the coupling
$\alpha(T)$ and the Debye mass $m_D(T)$, respectively. The fit-range was chosen
with respect to our discussion in Sec.~\ref{secshort}, {\em i.e.}
$rT\;\gsim\;0.8\;$-$\;1$, where we varied the lower fit limit within this range and
averaged over the resulting values.
The temperature dependent coupling $\alpha(T)$ defined here
will be discussed later. Our results for the screening mass, $m_D(T)/T$, are
summarized in Fig.~\ref{screenmass} as function of $T/T_c$ and are compared to
the results obtained in pure gauge theory \cite{Kaczmarek:2004gv}. The data
obtained from our $2$-flavor QCD calculations are somewhat larger than in
quenched QCD. Although we are not expecting perturbation theory to hold at
these small temperatures, this enhancement is in qualitative agreement with
leading order perturbation theory, {\em i.e.}
\begin{eqnarray}
\frac{m_D(T)}{T}&=&\left(1 + \frac{N_f}{6}\right)^{1/2}\;g(T)\;.\label{LOscreen}
\end{eqnarray}
However, using the $2$-loop formula (\ref{2loop}) to estimate the temperature
dependence of the coupling leads to significantly smaller values for $m_D/T$
even when setting the scale by $\mu=\pi$ which commonly is used as an upper
bound for the perturbative coupling. We therefore follow
\cite{Kaczmarek:2004gv,Kaczmarek:1999mm} and introduce a multiplicative
constant, $A$, {\em i.e.} we allow for a non-perturbative correction defined as
\begin{eqnarray}
\frac{m_D(T)}{T}&\equiv& A\;\left(1 + \frac{N_f}{6}\right)^{1/2}\;g_{2-loop}(T)\;,
\end{eqnarray}   
and fix this constants by best agreement with the non-perturbative data for
$m_D(T)/T$ at temperatures $T\;\gsim\;1.2$. Here the scale in the perturbative
coupling is fixed by $\mu=2\pi$. This analysis leads to $A=1.417(19)$ and is
shown as solid line with error band (dotted lines).  Similar results were
already reported in \cite{Kaczmarek:1999mm,Kaczmarek:2004gv} for screening
masses in pure gauge theory.  Using the same fit range, i.e.
$T=1.2T_c\;-\;4.1T_c$, for the quenched results, we obtain $A=1.515(17)$.  To
avoid here any confusion concerning $A$ we note that its value will crucially
depend on the temperature range used to determine it. When approaching the
perturbative high temperature limit, $A\to 1$ is expected.

\begin{figure}[tbp]
  \epsfig{file=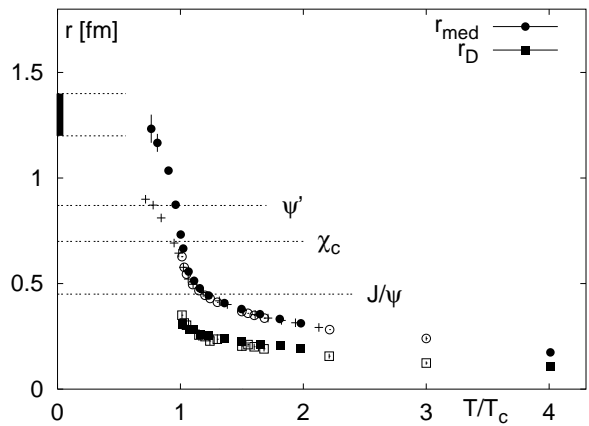,width=8.65cm}
\caption{
  The screening radius estimated from the inverse Debye mass, $r_D\equiv1/m_D$
  ($N_f$=0: open squares, $N_f$=2 filled squares), and the scale $r_{med}$
  ($N_f$=0: open circles, $N_f$=2: filled circles, $N_f$=3: crosses) defined in
  (\ref{rmed}) as function of $T/T_c$. The dotted line indicates the smallest
  separation available on lattices with temporal extent $N_\tau=4$. The
  horizontal lines gives the mean squared charge radii of some charmonium
  states, $J/\psi$, $\chi_c$ and $\psi\prime$ (see also \cite{Karsch:2005ex})
  and the band at the left frame shows the distance at which string breaking is
  expected in $2$-flavor QCD at $T=0$ and quark mass $m_\pi/m_\rho\simeq0.7$
  \cite{Pennanen:2000yk}.}
\label{screenradius}
\end{figure}

It is interesting to note here that the difference in $m_D/T$ apparent in
Fig.~\ref{screenmass} between $2$-flavor QCD and pure gauge theory disappears
when converting $m_D(T)$ to physical units. This is obvious from
Fig.~\ref{screenradius} which shows the Debye screening radius,
$r_D\equiv1/m_D$. In general $r_D$ is used to characterize the distance at
which medium modifications of the quark anti-quark interaction become dominant.
It often is used to describe the screening effects in phenomenological
inter-quark potentials at high temperatures. From perturbation theory one
expects that the screening radius will drop like $1/gT$. A definition of a
screening radius, however, will again depend on the ambiguities present in the
non-perturbative definition of a screening mass, $m_D(T)$. A different quantity
that characterizes the onset of medium effects, $r_{med}$, has already been
introduced in Sec.~\ref{secshort}; this quantity is also expected to drop like
$1/(gT)$ at high temperatures and could be considered to give an upper limit
for the screening radius \cite{Kaczmarek:2004gv}. In Fig.~\ref{screenradius} we
compare both length scales as function of temperature, $T/T_c$, and compare
them to the findings in quenched QCD \cite{Kaczmarek:2002mc,Kaczmarek:2004gv}.
It can be seen that in the temperature range analyzed here
$r_D(T)\;<\;r_{med}(T)$ and no or only little differences between the results
from quenched ($N_f$=0) and full ($N_f$=2,3) QCD could be identified. Again we
stress that in the perturbative high temperature limit differences are expected
to arise as expressed by Eq.~(\ref{LOscreen}).

It is important to realize that at distances well below $r_{med}$ medium
effects become suppressed and the color singlet free energy almost coincides
with the zero temperature heavy quark potential (see Fig.~\ref{fes}(a)). In
particular, the screening radius estimated from the inverse Debye mass
corresponds to distances which are only moderately larger than the smallest
distance available in our calculations (compare with the lower dotted line in
Fig.~\ref{onset}). In view of the almost temperature independent behavior of
the color singlet free energies at small distances (Fig.~\ref{fes}(a)) it could
be misleading to quantify the dominant screening length of the medium in terms
of $r_D\equiv1/m_D$. On the other hand the color averaged free energies show
already strong temperature dependence at distances similar to $r_D$ (see
Fig.~\ref{fes}(b)).

Following \cite{Karsch:2005ex} we also included in Fig.~\ref{screenradius} the
mean charge radii of the most prominent charmonium states, $J/\psi$, $\chi_c$
and $\psi\prime$, as horizontal lines. These lines characterize the averaged
separation $r$ which enters the effective potential in potential model
calculations. It thus is reasonable to expect that the temperature at which
these radii equal $r_{med}$ could give a rough estimate for the onset of
thermal effects in the charmonium states. It appears quite reasonable from this
view that $J/\psi$ indeed may survive the phase transition
\cite{Asakawa:2003re,Datta:2003ww}, while $\chi_c$ and $\psi\prime$ are
supposed to show significant thermal modifications at temperatures close to the
transition. Recent potential model calculations support this analysis
\cite{Wong:2004kn}. The wave functions for these states, however, will also
reach out to larger distances \cite{Jacobs:1986gv} and this estimate can only
be taken as a first indication for the relevant temperatures. Further details
on this issue including also bottomonium states have been given in
Ref.~\cite{Karsch:2005ex}. We will turn again to a discussion of thermal
modifications of quarkonium states in Ref.~\cite{pap2} using finite temperature
quark anti-quark energies.

\subsubsection{Temperature dependence of $\alpha_s$}
\begin{figure}[tbp]
  \epsfig{file=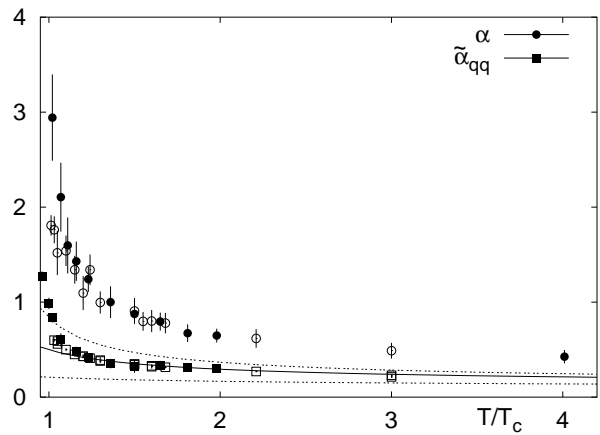,width=8.65cm}
\caption{
  The temperature dependence of the coupling $\alpha(T)$ as function of
  temperature (filled circles) from fits of the large distance behavior of free
  energies using Eq.~(\ref{screenfit}). We also show again the values of
  $\tilde{\alpha}_{qq}(T)$ defined as the maximum value of the running coupling
  $\alpha_{qq}(r,T)$ (filled squares) discussed in Sec.~\ref{couplatt} and the
  perturbative $2$-loop coupling, Eq.~(\ref{2loop}), with scales
  $\mu=\pi,...,4\pi$ (dashed lines). The open symbols indicate results from
  corresponding quenched QCD calculations \cite{Kaczmarek:2004gv}.}
\label{cTcomp}
\end{figure}
We finally discuss here the temperature dependence of the QCD coupling,
$\alpha(T)$, extracted from the fits used to determine also $m_D$, {\em i.e.}
from Eq.~(\ref{screenfit}). From fits of the free energies above deconfinement we
find the values shown in Fig.~\ref{cTcomp} as function of $T/T_c$ given by the
filled circles. We again show in this figure also the temperature dependent
coupling $\tilde{\alpha}_{qq}(T)$ introduced in Sec.~\ref{couplatt}. It can
clearly be seen that the values for both couplings are quite different,
$\tilde{\alpha}_{qq}(T)\;\gsim\;\alpha(T)$, at temperatures close but above
deconfinement while this difference rapidly decreases with increasing
temperature. This again demonstrates the ambiguity in defining the coupling in
the non-perturbative temperature range due to the different non-perturbative
contributions to the observable used for its definition
\cite{Kaczmarek:2004gv}. In fact, at temperatures close to the phase transition
temperature we find quite large values for $\alpha(T)$, {\em i.e.}
$\alpha(T)\simeq 2\; -\; 3$ in the vicinity of $T_c$, while it drops rapidly to
values smaller than unity, {\em i.e.} $\alpha(T)\;\lsim\;1$ already at
temperatures $T/T_c\;\gsim\;1.5$. A similar behavior was also found in
\cite{Kaczmarek:2004gv} for the coupling in pure gauge theory (open symbols).
In fact, no or only a marginal enhancement of the values calculated in full QCD
compared to the values in quenched QCD could be identified here at temperatures
$T\;\lsim\;1.5T_c$. We stress again that the large values for $\alpha(T)$ found
here should not be confused with the coupling that characterizes the short
distance Coulomb part of $F_1(r,T)$. The latter is almost temperature
independent at small distances and can to some extent be described by the zero
temperature coupling.

\subsection{String breaking below deconfinement}\label{stringbreaking}
We finally discuss the large distance properties of the free energies below
$T_c$. In contrast to the quark anti-quark free energy in quenched QCD where
the string between the quark anti-quark pair cannot break and the free energies are
linearly rising at large separations, in full QCD the string between two static
color charges can break due to the possibility of spontaneously generating
$q\bar q$-pairs from the vacuum. Therefore the quark anti-quark free energy
reaches a constant value also below $T_c$. In Fig.~\ref{fes} this behavior
is clearly seen.

\begin{figure}[tbp]
  \epsfig{file=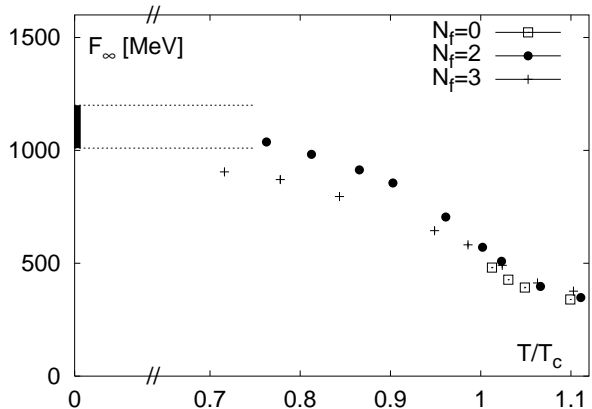,width=8.65cm}
\caption{The plateau value of the quark anti-quark free energy, $F_\infty(T)$,
  calculated in $2$-flavor QCD as function of $T/T_c$ at temperatures in the
  vicinity and below the phase transition. The dashed lines show the expected
  value at zero temperature, $V(r_{\text{breaking}})$, with
  $r_{\text{breaking}}\simeq1.2\sim1.4$ fm using quark mass
  $m_\pi/m_\rho\simeq0.7$ \cite{Pennanen:2000yk}. The open symbols
  ($T\;\gsim\;T_c$) correspond to $F_\infty(T)$ in quenched QCD (from
  Ref.~\cite{Kaczmarek:2002mc}) and the crosses to $3$-flavor QCD studies (from
  \cite{Petreczky:2004pz}). The relative normalization of the corresponding
  $V(r)$ used for renormalization of $F_1(r\to\infty,T)$ in quenched and full
  QCD is such that the Cornell parameterization of $V(r)$ does not contain any
  constant at large distances. }
\label{string1}
\end{figure}
The distances at which the quark anti-quark free energies approach an almost
constant value move to smaller separations at higher temperatures. This can
also be seen from the temperature dependence of $r_{med}$ in Fig.~\ref{onset}
at temperature below $T_c$. By construction $r_{med}$ describes a distance
which can be used to estimate a lower limit for the distance where the string
breaking will set in. An estimate of the string breaking radius at $T=0$ can be
calculated from the lightest heavy-light meson,
$r_{\text{breaking}}\simeq1.2-1.4$ fm \cite{Pennanen:2000yk} and is shown on
the left side in Fig.~\ref{onset} within the dotted band. It can be seen that
$r_{med}$ in $2$-flavor QCD does indeed approach such values at temperatures
$T\;\lsim\;0.8T_c$. This suggests that the dependence on temperature in
$2$-flavor QCD is small below the smallest temperature analyzed here,
$0.76T_c$. This can also be seen from the behavior of $F_\infty(T)$ shown in
Fig.~\ref{string1} (see also Fig.~\ref{fes}(a)) compared to the value commonly
expected at $T=0$. We use $V(r_{\text{breaking}})\simeq1000-1200$ MeV as
reference to the zero temperature string breaking energy with quark mass
$m_\pi/m_\rho\simeq0.7$.  This estimate is shown on the left side in
Fig.~\ref{string1} as the dotted band. A similar behavior is expected for the
free energies in $3$-flavor QCD and smaller quark mass,
$m_\pi/m_\rho\simeq0.4$. As seen also in Fig.~\ref{string1} the values for
$F_\infty(T)$ are smaller than in $2$-flavor QCD and larger quark mass.  This
may indicate that string breaking sets in at smaller distances for smaller
quark masses. However, in \cite{Karsch:2000kv} no mass dependence (in the color
averaged free energies) was observed below the quark mass analyzed by us
($m/T$=0.4). At present it is, however, difficult to judge whether the
differences seen for $2$- and $3$-flavor QCD for $T/T_c<1$ are due to quark
mass or flavor dependence of the string breaking. Although $F_\infty(T)$ still
is close to $V(r_{\text{breaking}})$ at $T\sim0.8T_c$, it rapidly drops to
about half of this value in the vicinity of the phase transition,
$F_\infty(T_c)\simeq575$ MeV. This value is almost the same in $2$- and
$3$-flavor QCD; we find $F_\infty^{N_f=2}(T_c)\simeq575(15)$ MeV and
$F_\infty^{N_f=3}(T_c)\simeq548(20)$ MeV. It is interesting to note that also
the values of $F_\infty(T)$ in quenched QCD ($N_f$=0) approach a similar value
at temperatures just above $T_c$. We find $F_\infty(T_c^+)\simeq481(4)$ MeV
where $T_c^+\equiv1.02T_c$ denotes the closest temperature above $T_c$ analyzed
in quenched QCD. Of course, the value for $F_\infty(T_c^+)$ will increase when
going to temperatures even closer to $T_c$. The flavor and quark mass
dependence of $F_\infty(T)$ including also higher temperatures will be
discussed in more detail in Ref.~\cite{pap2}.

\section{Summary and Conclusions}\label{seccon}
Our analysis of the zero temperature heavy quark potential, $V(r)$, calculated
in $2$-flavor lattice QCD using large Wilson loops \cite{Karsch:2000kv} shows
no signal for string breaking at distances below $1.3$ fm. This is quite
consistent with earlier findings \cite{Bernard:2001tz,Pennanen:2000yk}. The
$r$-dependence of $V(r)$ becomes comparable to the potential from the bosonic
string picture already at distances larger than $0.4$ fm. Similar findings have
also been reported in lattice studies of the potential in quenched QCD
\cite{Necco:2001xg,Luscher:2002qv}. At those distances, $0.4$
fm$\;\lsim\;r\;\lsim\;1.5$ fm, we find no or only little differences between
lattice data for the potential in quenched ($N_f$=0) given in
Ref.~\cite{Necco:2001xg} and full ($N_f$=2) QCD. At smaller distances, however,
deviations from the large distance Coulomb term predicted by the string
picture, $\alpha_{\text{str}}\simeq0.196$, are found here when performing best
fit analysis with a free Cornell Ansatz. We find $\alpha\simeq0.212(3)$ which
could describe the data down to $r\;\gsim\;0.1$ fm. By analyzing the coupling
in the $qq$-scheme defined through the force, $dV(r)/dr$, small enhancement
compared to the coupling in quenched QCD is found for $r\;\lsim\;0.4$ fm. At
distances substantially smaller than $0.1$ fm the logarithmic weakening of the
coupling enters and will dominate the $r$-dependence of $V(r)$. The observed
running of the coupling may already signal the onset of the short distance
perturbative regime. This is also evident from quenched QCD lattice studies of
$V(r)$ \cite{Necco:2001gh}.

The running coupling at finite temperature defined in the $qq$-scheme using the
derivative of the color singlet quark anti-quark free energy, $dF_1(r,T)/dr$,
shows only little qualitative and quantitative differences when changing from
pure gauge \cite{Kaczmarek:2004gv} to full QCD at temperatures well above
deconfinement. Again, at small distances the running coupling is controlled by
distance and becomes comparable to $\alpha_{qq}(r)$ at zero temperature. The
properties of $\alpha_{qq}(r,T)$ at temperatures in the vicinity of the phase
transition are to large extent controlled by the confinement signal at zero
temperature. A clear separation of the different effects usually described by
the concepts of color screening ($T\;\gsim\;T_c$) and the concept of string
breaking ($T\;\lsim\;T_c$) is difficult in the crossover region. Remnants of
the confinement part of the QCD forces may in parts dominate the
non-perturbative properties of the QCD plasma at temperatures only moderately
larger than $T_c$. This supports similar findings in recent studies of the
quark anti-quark free energies in quenched QCD \cite{Kaczmarek:2004gv}.

The properties of the quark anti-quark free energy and the coupling at small
distances thus again allow for non-perturbative renormalization of the free
energy and Polyakov loop \cite{Kaczmarek:2002mc}. The crossover from
confinement to deconfinement is clearly signaled by the Polyakov loop through a
rapid increase at temperatures close to $T_c$. String breaking dominates the
quark anti-quark free energies at temperatures well below deconfinement in all
color channels leading to finite values of the Polyakov loop. The string
breaking energy, $F_\infty(T)$, and the distance where string breaking sets in,
are decreasing with increasing temperatures. The plateau value $F_\infty(T)$
approaches about $95\%$ of the value one usually estimates at zero temperature,
$V(r_{\text{breaking}})\simeq1.1$ GeV \cite{Pennanen:2000yk,Bernard:2001tz},
already for $T\;\simeq\;0.8T_c$. We thus expect that the change in quark
anti-quark free energies is only small when going to smaller temperatures and
the quark anti-quark free energy, $F_1(r,T)$, will show only small differences
from the heavy quark potential at $T=0$, $V(r)$. Significant thermal
modifications on heavy quark bound states can thus be expected only for
temperatures above $0.8T_c$. Our analysis of $r_{med}$ suggests indeed a
qualitative similar behavior for the free energies in $3$-flavor QCD. This can
also be seen from the behavior of $r_{med}$ shown in Fig.~\ref{screenradius}.

At temperatures well above the (pseudo-) critical temperature, {\em i.e.}
$1.2\;\lsim\;T/T_c\;\lsim\;4$, no or only little qualitative differences in the
thermal properties of the quark anti-quark free energies calculated in quenched
($N_f$=0) and full ($N_f$=2,3) QCD could be established here when converting
the observables to physical units. Color screening clearly dominates the quark
anti-quark free energy at large distances and screening masses, which are
non-perturbatively determined from the exponential fall-off of the color
singlet free energies, could be extracted (for $N_f$=2). In accordance with
earlier findings in quenched QCD \cite{Kaczmarek:1999mm,Kaczmarek:2004gv} we
find substantially larger values for the screening masses than given by leading
order perturbation theory. The values of the screening masses, $m_D(T)$, again
show only marginal differences as function of $T/T_c$ compared to the values
found in quenched QCD (see also Fig.~\ref{screenradius}). The large screening
mass defines a rather small screening radius, $r_D\equiv1/m_D$, which refers to
a length scale where the singlet free energy shows almost no deviations from
the heavy quark potential at zero temperature. It thus might be misleading to
quantify the length scale of the QCD plasma where temperature effects dominate
thermal modifications on heavy quark bound states with the observable
$r_D\equiv1/m_D$ in the non-perturbative temperature regime close but above
$T_c$. On the other hand the color averaged free energies show indeed strong
temperature dependence at distances which could be characterized by $1/m_D$. In
view of color changing processes as a mechanism for direct quarkonium
dissociation \cite{Kharzeev:1994pz} the discussion of the color averaged free
energy could become important.

We have also compared $r_D$ and $r_{med}$ in Fig.~\ref{screenradius} to the
expected mean squared charge radii of some charmonium states. It is reasonable
that the temperatures at which these radii equal $r_{med}$ give a first
indication of the temperature at which thermal modifications become important
in the charmonium states. It appears thus quite reasonable that $J/\psi$ will
survive the transition while $\chi_c$ and $\psi\prime$ are expected to show
strong thermal effects at temperatures in the vicinity of the transition and
this may support recent findings
\cite{Wong:2004kn,Asakawa:2003re,Petreczky:2003js}. Of course the wave
functions of these states will also reach out to larger distances and thus our
analysis can only be taken as a first indication of the relevant temperatures.
We will turn back to this issue in Ref.~\cite{pap2}. The analysis of bound
states using, for instance, the Schr\"odinger equation will do better in this
respect. It can, however, clearly be seen from Fig.~\ref{screenradius} that
although $r_{med}(T_c)\simeq0.7$ fm is approached almost in common for
$N_f$=0,2,3, it falls apart for $N_f$=2,3 at smaller temperatures. It thus
could be difficult to determine suppression patterns from free energies for
quarkonium states which are substantially larger than $0.7$ fm independently
from $N_f$ and/or finite quark mass.

The analysis presented here has been performed for a single quark mass value
that corresponds to a pion mass of about $770$ MeV ($m_\pi/m_\rho\simeq0.7$).
In Ref.~\cite{Karsch:2000kv}, however, no major quark mass effects were visible
in color averaged free energies below this quark mass value. The comparisons of
$r_{med}$ and $F_\infty(T)$ calculated in $2$-flavor ($m_\pi/m_\rho\simeq0.7$)
with results calculated in $3$-flavor QCD ($m_\pi/m_\rho\simeq0.4$
\cite{Peterpriv}) supports this property. While at temperatures above
deconfinement no or only little differences in this observable can be
identified, at temperatures below $T_c$ differences can be seen.  To what
extend these are due to the smaller quark masses used in the $3$-flavor case or
whether these differences reflect a flavor dependence of the string breaking
distance requires further investigation.  
The present analysis was carried out on one lattice size ($16^3\times 4$)
and therefore performing an extrapolation to the continuum limit could not be
done with the current data. However the analysis of the quenched free
energies \cite{Kaczmarek:2002mc,Kaczmarek:2004gv},where no major differences
between the $N_\tau=4$ and $N_\tau=8$ results were visible, and the use of
improved actions suggests that
cut-off effects might be small.
Despite these uncertainties and the
fact that parts of our comparisons to results from quenched QCD are on a
qualitative level, we find quite important information for the study of heavy
quark bound states in the QCD plasma phase. At temperatures well above $T_c$,
{\em i.e.}  $1.2\;\lsim\;T/T_c\;\lsim\;4$, no or only little differences appear
between results calculated in quenched and QCD. This might suggest that using
thermal parameters extracted from free or internal energy in quenched QCD as
input for model calculations of heavy quark bound states
\cite{Shuryak:2004tx,Wong:2004kn} is a reasonable approximation. Furthermore
this also supports the investigation of heavy quarkonia in quenched lattice QCD
calculations using the analysis of spectral functions
\cite{Datta:2003ww,Asakawa:2003re,Asakawa:2002xj}. On the other hand, however,
most of our $2$- and $3$-flavor QCD results differ from quenched calculations
at temperatures in the vicinity and below the phase transition. Due to these
qualitative differences, results from quenched QCD could make a discussion of
possible signals for the quark gluon plasma production in heavy ion collision
experiments complicated when temperatures and/or densities close to the
transition become important.

\begin{acknowledgments}
  We thank the Bielefeld-Swansea collaboration for providing us their
  configurations with special thanks to S. Ejiri. We would like to thank E.
  Laermann and F. Karsch for many fruitful discussions. F.Z. thanks P.
  Petreczky for his continuous support. We thank K. Petrov and P. Petreczky for
  sending us the data of Ref.~\cite{Petreczky:2004pz}. This work has partly
  been supported by DFG under grant FOR 339/2-1 and by BMBF under grant
  No.06BI102 and partly by contract DE-AC02-98CH10886 with the U.S. Department
  of Energy. At an early stage of this work F.Z. has been supported through a
  stipend of the DFG funded graduate school GRK881. Some of the results
  discussed in this article were already presented in proceeding contributions
  \cite{Kaczmarek:2003ph,Kaczmarek:2005uv,Kaczmarek:2005uw}.
\end{acknowledgments}

\bibliographystyle{h-physrev3} \bibliography{paper}

\end{document}